\documentclass[aps,prb,a4paper,twocolumn,showpacs,showkeys,fleqn,floatfix,superscriptaddress]{revtex4-2}

\usepackage{graphicx}
\usepackage{color}
\usepackage{amsmath}
\usepackage{multirow}
\usepackage{dcolumn}

\newcommand{\nabok}[1]{{#1}Cu$_7$(TeO$_4$)(SO$_4$)$_5$Cl}
\newcommand{\kagome}{kagom\'{e}}

\newcommand{\abs}[1]{\left|#1\right|}

\begin{document}

\title{High-frequency dielectric anomalies in a highly frustrated square \kagome{} lattice nabokoite family compounds \nabok{A}{} (A=Na, K, Rb, Cs). }

\author{Ya.V.~Rebrov}
\email{yavrebrov@edu.hse.ru}
\affiliation{P.L.~Kapitza Institute for Physical Problems, RAS, 119334, Moscow, Russia}
\affiliation{National Research University  ``HSE'', 109028, Moscow, Russia}

\author{V.N.~Glazkov}
\email{glazkov@kapitza.ras.ru}
\affiliation{P.L.~Kapitza Institute for Physical Problems, RAS, 119334, Moscow, Russia}
\affiliation{National Research University  ``HSE'', 109028, Moscow, Russia}

\author{A.F.~Murtazoev}
\affiliation{M.V.~Lomonosov Moscow State University, Moscow 119991, Russia}

\author{V.A.~Dolgikh}
\affiliation{M.V.~Lomonosov Moscow State University, Moscow 119991, Russia}

\author{P.S.~Berdonosov}
\affiliation{M.V.~Lomonosov Moscow State University, Moscow 119991, Russia}
\affiliation{National University of Science and Technology ``MISiS'', Moscow 119049, Russia}

\begin{abstract}
Nabokoite family compounds \nabok{A}{} (A=Na, K, Cs, Rb) are  candidates for the experimental realization of highly-frustrated 2D square \kagome{} lattice (SKL). Their magnetic subsystem includes SKL layers decorated by additional copper ions. All members of this family are characterized by quite high Curie-Weiss temperatures ($\sim 80-200$~K), but magnetic ordering was reported only for Na and K compounds at a much lower temperatures below 4~K. We report here results of the study of high-frequency ($\sim 10$~GHz) dielectric properties of this family of compounds. Our study revealed presence of the strong dielectric anomaly both in the real and imaginary parts of high-frequency dielectric permittivity for Na and K compounds  approx. 100 and 26~K, correspondingly, presumably related to antiferroelectric ordering. Additionally, much weaker anomalies were observed at approx. 5~K  indicating possible interplay of magnetic and lattice degrees of freedom. We discuss possible relation between the structure rearrangements accompanying dielectric anomalies and a delayed magnetic ordering in the nabokoite family compounds.
\end{abstract}

\maketitle

\section{Introduction}
Frustrated magnets remains in the focus of magnetism during recent decades \cite{anderson1973,anderson1978,ramirez1994,lacroix2011}. Common feature of these magnets is presence of the competing interactions, which, by opposing each other, prevent formation of a conventional magnetic ordering at temperatures of the scale of Curie-Weiss temperature. Varieties of geometrically frustrated systems were discovered and studied both experimentally and theoretically supplying new physical concepts and phenomena, such as various spin-liquid states \cite{balentz2010}, fragmentation of excitations \cite{lhotel2020} or a Kitaev model \cite{kitaev2006}. Phase diagrams of geometrically frustrated magnets are also a subject of interest since minute changes caused by external field and hydrostatic or chemical pressure can shift the fine balance of the competing interaction in favor of different ground states.

Kagom\'{e} lattice constructed from the corner-sharing triangles with hexagonal voids is one of the archetypical models of geometric frustration, which is studied since being introduced in 1951 by I.Sy\^{o}zi \cite{syozi1951,kano1953}. Despite of long history behind this problem, there is a lasting  theoretical discussions on whether the spin-liquid state of the Heisenberg \kagome{} magnet is gapped or gapless\cite{Depenbrock2012,liao2017,mei2017,lauchli2019}. Experimental realizations of Heisenberg \kagome{} magnets remains quite scarce, herbertsmithite ZnCu$_3$(OH)$_6$Cl$_2$ being the most promising (albeit containing 5-10\% of intrinsic defects in the form of ``extra'' copper ions taking over inter-planar positions of zinc ions)  \cite{olariu2008,wang2021}.

Extension of the \kagome{} model, called square \kagome{} lattice (SKL), was proposed in 2001 \cite{Siddharthan2001} as a 2D lattice of corner-sharing triangles with alternating square and octagonal voids. Theoretical considerations predict various possible ground states for SKL antiferromagnet depending on the ratio of  exchange coupling parameters \cite{richter2009,rousochatzakis2013,morita2018,Lugan2019,richter2023,richter2023noncoplanar}. Detailed numeric simulations demonstrate that a spin-liquid state with gapped triplet excitations and singlet excitations within the triplet gap is formed in equilateral Heisenberg SKL model  \cite{Richter2022}. Experimental realizations of the square \kagome{} lattice were found in some families of minerals and their synthetic siblings \cite{fuji2020,liu2022,Murtazoev2023}. Obtained experimental results for different materials can be interpreted in favor of both gapless \cite{fuji2020} and gapped \cite{kolhoz} spin-liquid state, this controversial difference most likely being caused by the effects of additional interactions in real compounds.

In the present study we focus on the nabokoite family compounds \nabok{A}{} (A=Na, K, Rb, Cs). These compounds were reported recently as a spin-liquid state candidates \cite{Murtazoev2023,kolhoz}. The reported Curie-Weiss temperature for these compounds is of antiferromagnetic sign with magnitude of 80--200~K, however, magnetic ordering is reported for K- and Na-nabokoites only at the temperatures below 4~K \cite{Murtazoev2023,kolhoz}.  I.e., a broad temperature window for a strongly correlated spin-liquid-like state is unambiguously observed in the nabokoite family compounds in earlier studies. Specific of the nabokoite family compounds compared to ideal SKL model is that the square \kagome{} lattice there is decorated with additional inter-planar copper ions. Such a geometry of exchange bonds (``decorated SKL'') was not considered theoretically so far.

Magnetic ordering in nabokoites is of separate interest. Only Na- and K-nabokoites were reported to order at low temperatures \cite{Murtazoev2023,kolhoz}. Electron spin resonance data \cite{kolhoz} provides evidences for noncollinear ordering in \nabok{K}{}, possibility of non-trivial ordering in  SKL models including different couplings was also considered theoretically \cite{morita2018,Lugan2019,richter2023noncoplanar}. Static dielectric anomaly was reported for K nabokoite \cite{kolhoz} at 25~K, probably indicating antiferroelectric transition, which, being accompanied by some lattice deformation, can lift frustration of some of the exchange bonds leading thus towards magnetic ordering at lower temperature. This finding additionally stimulates interest to nabokoite family because of possible interplay between electric and magnetic degrees of freedom, which is prerequisite to multiferroic-like behavior.

Here we report results on the high-frequency (9-14~GHz) dielectric anomalies sought in all members of nabokoite family \nabok{A} (A=Na, K, Rb, Cs). We have found that K and Na compounds demonstrate strong dielectric anomalies in both real and imaginary parts of dielectric permittivity at approximately 25 and 100~K, correspondingly, while Rb and Cs compounds do not demonstrate any anomaly of the similar strength from 1.7 to 300~K. Nevertheless, much weaker dielectric response was observed in all members of nabokoite family at different temperatures around 5~K. We argue that lattice deformations accompanying observed dielectric anomaly can be the key to understanding of the magnetic ordering in nabokoites.

\section{Experimental details and samples}

\subsection{Samples preparation}
We used polycrystalline samples of \nabok{A} from the same growth parties as were used in the earlier structural, thermodynamic and magnetic resonance studies \cite{Murtazoev2023,kolhoz}. Polycrystalline sample preparation is described in details in \cite{Murtazoev2023}.

Samples were conditioned for the high-frequency experiment as follows (see also Figure~\ref{fig:setup}): thin-wall plastic straw (3.2 mm outer diameter) was sealed at one end with the small PTFE 'cup', small amount ($\approx 4$~mg) of nabokoite powder was placed into this 'cup' and soaked with ethanol-dissolved glue (``BF-2'' trademark). Once the ethanol evaporates and glue dries the sample takes form of the thin disks (diameter 3~mm, thickness about 0.2~mm). Empty sample-holder (containing approximately the same quantity of plastic straw, PTFE and glue) was prepared for reference measurements. Larger samples (15-20~mg) were prepared for some of the compounds for qualitative experiments in search of weak dielectric anomalies.

\subsection{Setup for high-frequency measurements and data processing}

\begin{figure*}[th]
\centering
     \includegraphics[width=0.8\textwidth]{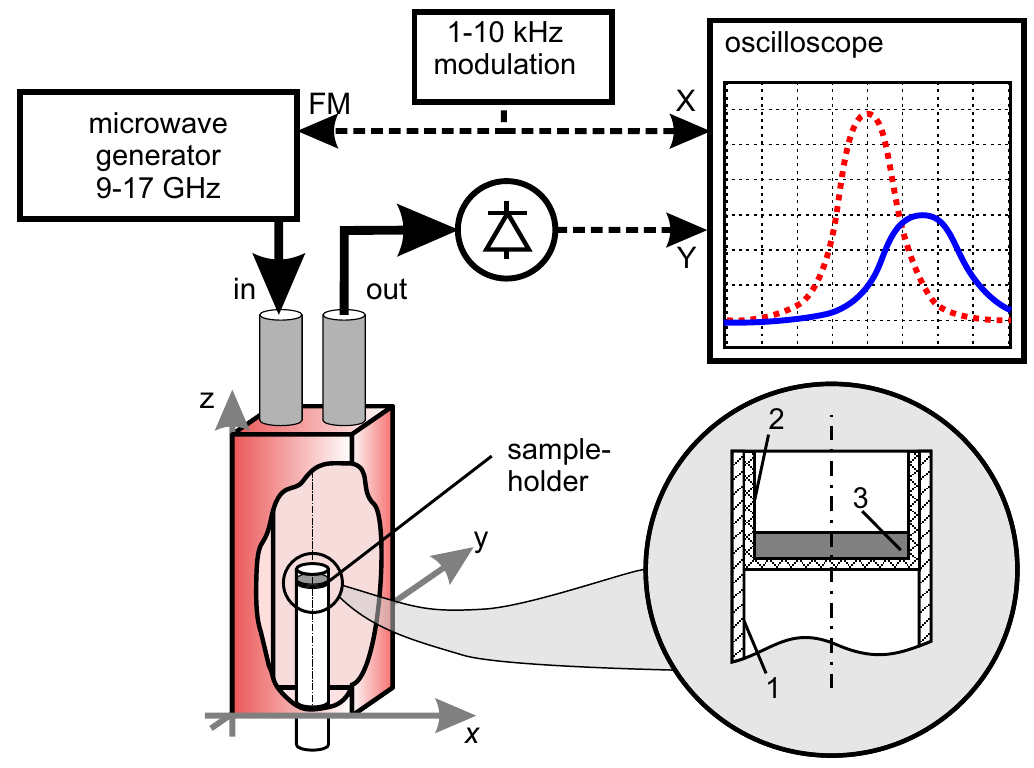}
    \caption{Schematic drawing of the experimental setup used for high frequency experiments. Sample is placed in the center of rectangular cavity. Frequency-modulated microwave generator sweep over the cavity resonance curve allowing to follow shift of the eigenfrequency and bandwidth with temperature. Oscilloscope 'screen' shows reference curve (dashed curve) and shifted and broadened (solid curve) cavity resonance curves. Expanded view demonstrate sample and sample-holder details: (1) thin wall plastic straw, (2) PTFE 'cup', (3) powder sample.}
    \label{fig:setup}
\end{figure*}

We used resonator technique to measure high-frequency dielectric properties of the sample (see Figure~\ref{fig:setup}). We have measured transmission of microwave signal through the cavity with the sample at frequencies $9.6, 11.6$ and $14.3$~GHz corresponding to TE101, TE102 and TE103 eigenmodes of the rectangular resonator. Sample was placed in the center of the rectangular cavity (cavity dimensions $\simeq 7\times 17\times 40$~mm$^3$), which is  the beam of the high-frequency electric field for  TE101 and TE103 eigenmodes and  the beam of the high-frequency magnetic field for TE102 eigenmode (see Appendix for the brief summary on microwave technique). Sample size was small compared to the microwave wavelength ($\lambda\simeq 3$~cm), sample plane was parallel to the high-frequency electric field at the sample location (for  TE101 and TE103 eigenmodes). Comparison of the results obtained at TE101, TE102 and TE103 eigenmodes allows to differ unambiguously high-frequency dielectric and magnetic effects in the sample.

The microwave cavity was surrounded by a hermetically sealed shell filled with a small amount of heat exchange helium gas. The shell was submerged into liquid helium or liquid nitrogen to reach low temperatures. Liquid helium bath cryostat was used at 1.7--100 K temperature range, liquid nitrogen bath cryostat was used at 77-300 K temperature range. The cavity temperature change during the experiment (cavity temperature is the same as the sample temperature) causes some change of its' eigenfrequency and bandwidth due to  the effects of thermal contraction and to the temperature-dependent cavity walls resistance. These effect were accounted for by performing reference experiments with the empty sample-holder.

The real part of sample's dielectric permittivity $\varepsilon'\neq 1$  causes shift of the cavity eigenfrequency at the given temperature: $$\delta\omega=\omega_0^{(\textrm{s})}-\omega_0^{(\textrm{e})}$$
(here and further on indices $(\textrm{s})$ and $(\textrm{e})$ denote measurements with the sample-loaded resonator and with the empty sample-holder, correspondingly), while the dielectric losses described by $\varepsilon''$ affect cavity Q-factor $$Q=\omega_0/(\Delta\omega)$$ (here $\Delta\omega$ is the cavity bandwidth, full width at half maximum amplitude). Quantities $\omega_0(T)$ and $\Delta\omega(T)$ are obtained from Lorentzian fits of the cavity transmission curve measured via frequency modulation of microwave generator. Typical variation of the cavity eigenfrequency over the full temperature range ($1.7...300$~K) was about 40--50~MHz, typical bandwidth was 3--10~MHz at different eigenmodes ($Q\simeq 1000...4000$).

We have had no technical means to measure absolute value of the microwave frequency with the accuracy better than 1~MHz to compare directly $\omega_0(T)$ results for the sample-loaded cavity and for the empty sample-holder. Frequency modulation technique, on the contrary, allows to follow variation of the cavity eigenfrequency in the given experiment with sufficient accuracy. Thus, we have measured variation of the cavity eigenfrequency compared to its lowest temperature value $$v(T)=\omega_0(T)-\omega_0(1.7\textrm{~K})$$ and contrast $v^{(\textrm{s})}(T)$ vs. $v^{(\textrm{e})}(T)$. We have found, that (i) $v(T)$ dependence is reproducible for the given sample and (ii)  $v(300\textrm{~K})$ is the same within the error margin for all samples and for empty sample-holder. This allows to calculate frequency shift $\delta\omega$ introduced above as the difference of $v(T)$ curves measured with the sample-loaded resonator and with the empty sample-holder: $$\delta\omega(T)=v^{(\textrm{s})}(T)-v^{(\textrm{e})}(T).$$
 This approach ignores possible temperature-independent part of the high-frequency dielectric permittivity allowing to determine only the ``anomalous'' part.

The real part of dielectric permittivity (its ``anomalous'' part) $\varepsilon'$ can be found from the cavity eigenfrequency shift $\delta\omega$:
\begin{equation}
    \varepsilon' = 1 - \frac{V^{(\textrm{res})}}{2V^{(\textrm{s})}}\cdot \frac{\delta \omega}{\omega_0} \label{eq.eps'}
\end{equation}
\noindent and the imaginary part $\varepsilon''$ can be determined from the change of the cavity Q-factor:
\begin{equation}
     \varepsilon'' = \frac{V^{(\textrm{res})}}{4 V^{(\textrm{s})}}\cdot \left(\frac{1}{Q^{(\textrm{s})}} - \frac{1}{Q^{(\textrm{e})}}\right)\label{eq.eps''}
\end{equation}

\noindent  here $V^{(\textrm{s})}$ and $V^{(\textrm{res})}$ are the volumes of the sample and that of the resonator, correspondingly, sample volume was determined from its mass using density computed from the structural data \cite{Murtazoev2023}. Equations above assume that the sample size is small compared to the microwave wavelength and that the sample is exactly at the beam of the high-frequency electric field for TE$10k$ eigenmode. Details of these equations derivations are given in the Appendix.

\subsection{Error analysis}

Our main experimental results are (i) qualitative detection of the high-frequency dielectric anomalies in nabokoite compounds and (ii) quantitative determination of real and imaginary parts of the high-frequency dielectric permittivity for strong dielectric anomalies in K and Na nabokoites.

Qualitative detection of the high-frequency dielectric anomalies is double-controlled by the measurements' reproducibility  and by the observation of simultaneous response in real ($\varepsilon'$)  and imaginary ($\varepsilon''$) parts of the high-frequency dielelctric permittivity. Dielectric origin of the observed anomaly was confirmed by its observation only in the TE101 and TE103 modes experiments, when the sample is in the beam of the high-frequency electric field, and by the absence of similar anomaly in the experiments with TE102 mode, when the sample is in the node of the electric field. This makes qualitative observation of the dielectric anomalies a well established experimental result.

Quantitative calculation of $\varepsilon'$, $\varepsilon''$ via equations (\ref{eq.eps'}), (\ref{eq.eps''}) is subject to several potential errors. Main error contributions are due to the slight sample misplacement from the geometrical center of the cavity and to the uncontrollable slight alternation of microwave transmission line on each assembly of experimental setup. Additionally we have to underline, that our experiments are performed on powder sample, which adds averaging over particles orientations. Reasonable estimate of systematic error caused by the sample misplacement is the difference of the calculated dielectric permittivities in TE101 and TE103 modes experiments: the peak amplitude of the dielectric permittivity curves can differ by up to 20\%. Random errors, which are due to various noises etc., were estimated by repeating measurements several times in the same setup which yielded dielectric permittivity curves differing by approximately 5\% of the peak value.

\section{Experimental results}
\begin{figure*}[th]
\centering
     \includegraphics[width=\textwidth]{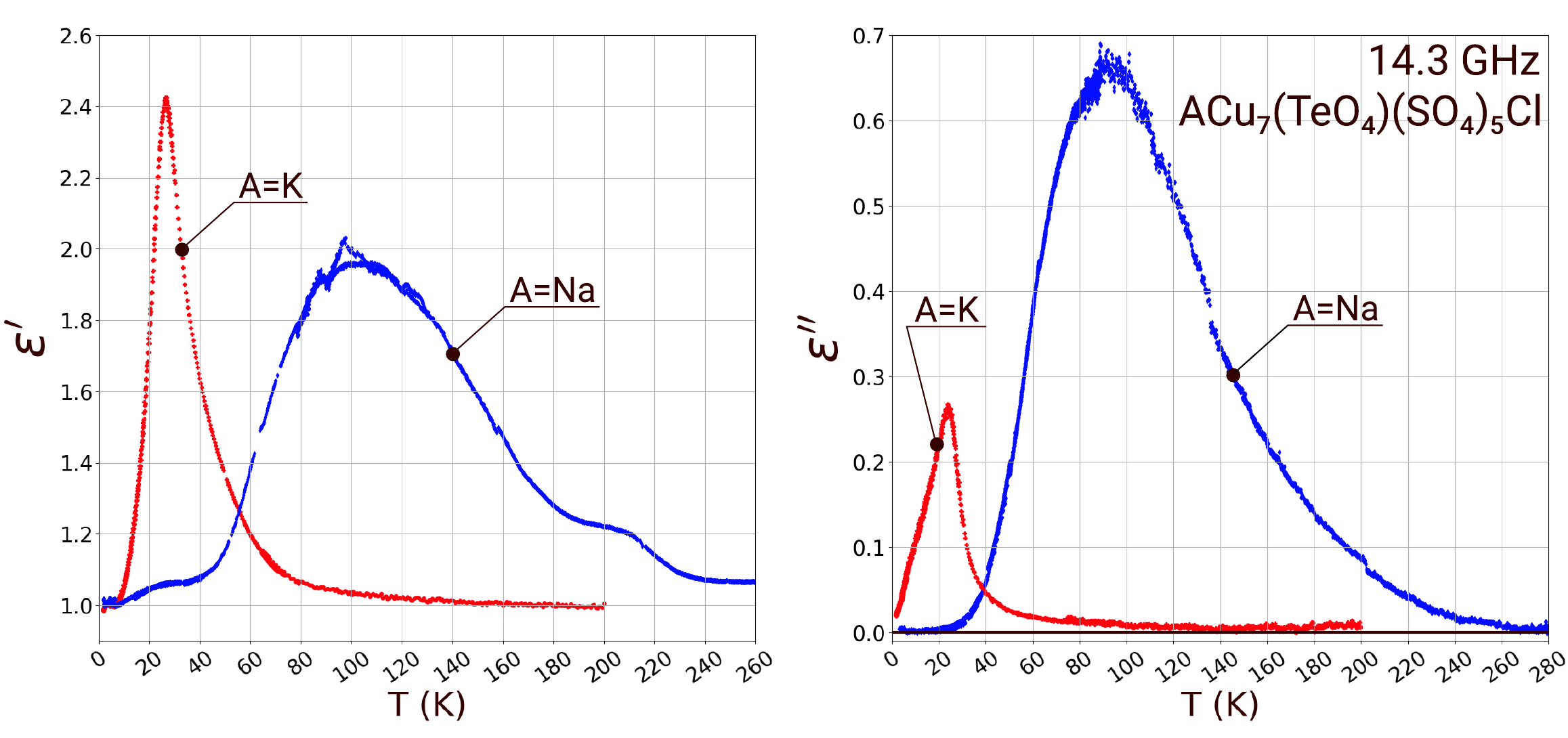}
    \caption{Temperature dependence of the real part $\varepsilon'$ (left graph) and imaginary part $\varepsilon''$ (right graph) of the high-frequency dielectric permittivity at 14.3~GHz. Red marks correspond to \nabok{K}, and blue marks to \nabok{Na}.}
    \label{fig:eps'-eps''}
\end{figure*}

\begin{figure*}[ht]
 \centering
\includegraphics[width=0.7\textwidth]{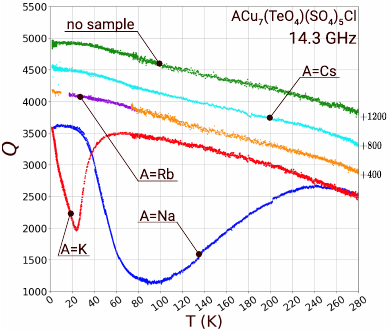}
 \caption{Temperature dependence of the resonator Q-factor at 14.3~GHz (TE103 eigenmode)  in experiments with \nabok{A} (A=Na, K, Rb, Cs) and reference curve for empty resonator. The data for Rb-, Cs- nabokoites and for empty resonator  are shifted by the values indicated on the graph for better representation. Samples masses: 4.5~mg for K-nabokoite, 4.0~mg for Na-nabokoite, 4.2~mg for Rb-nabokoite at $15\textrm{ K}<T<70\textrm{ K}$, 21~mg for Rb-nabokoite at other temperatures, 3.5~mg for Cs-nabokoite.}
 \label{fig:quality}
\end{figure*}

\begin{figure*}[ht]
 \centering
\includegraphics[width=0.7\textwidth]{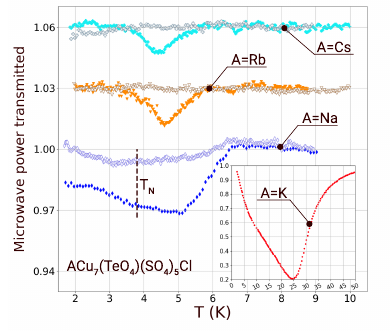}
 \caption{Low-temperature features in dielectric response of  \nabok{A} (A=Na, K, Rb, Cs). Main panel: temperature dependences of the microwave power transmitted at 14.3~GHz (filled symbols, TE103 eigenmode, sample in the beam of the high-frequency electric field) and at 11.7~GHz (open symbols, H$_{102}$ eigenmode, sample in the beam of the high-frequency magnetic field) for A=Na, Rb, Cs. Data for Rb and Cs compound are shifted for better presentation, vertical dashed line for Na compound marks known N\'{e}el temperature \cite{Murtazoev2023}. External field of 400~Oe is applied to suppress superconductivity in PbSn-soldered resonator. Inset: temperature dependence of the  microwave power transmitted for K-nabokoite in the temperature range including the main anomaly.  Samples masses: 4.5~mg for K-nabokoite, 14.5~mg for Na-nabokoite, 21~mg for Rb-nabokoite and 14.4~mg for Cs-nabokoite.}
 \label{fig:lt}
\end{figure*}

The  measurements on all four members of \nabok{A} (A=Na, K, Rb, Cs) nabokoite family were carried out at frequencies 9.6, 11.6 and 14.3~GHz (cavity eigenmodes TE101, TE102 and TE103, correspondingly) in the temperature range from 1.7  to 300~K.

We have observed strong anomalies of dielectric origin in the case of potassium and sodium nabokoites. Temperature dependences of real ($\varepsilon'$) and imaginary ($\varepsilon''$) parts of the high-frequency  dielectric permittivity extracted from the observed anomalies via equations (\ref{eq.eps'}), (\ref{eq.eps''}) are shown in Figure~\ref{fig:eps'-eps''}.

For \nabok{K}{} the maximum of the real part of dielectric permittivity $\varepsilon'$ is observed at $(26.5\pm 0.5)$~K and the peak of imaginary part $\varepsilon''$ at ($24\pm 1)$~K. For \nabok{Na}{} the maximum of the real part $\varepsilon'$ is observed at $(100\pm 5)$~K and the peak of the imaginary part $\varepsilon''$ at $(90\pm10)$~K. Position of the peak of the imaginary part for both compounds is slightly shifted to lower temperatures.
For the K-nabokoite sample it was verified that an external magnetic field up to 4~T does not affect the shape or position of the observed anomaly. Temperature of the observed high-frequency dielectric anomaly in \nabok{K}{} practically coincides with the position of static dielectric permittivity $\varepsilon(T)$ anomaly at 25.4~K \cite{kolhoz}.

Temperature dependences of real and imaginary parts of the dielectric permittivity measured at 9.6~GHz and at 14.3~GHz coincides within the error margin. Qualitatively the same behavior (strong microwave absorption close to the temperature of dielectric anomaly) was earlier observed at Q-band (30~GHz range) electron spin resonance experiments on \nabok{K}{} \cite{kolhoz}. All these observations suggest that the observed dielectric anomaly has a non-resonant character.

No dielectric anomalies comparable in magnitude to those of the K- or Na-nabokoites were observed over the entire temperature range for \nabok{Rb}{} and \nabok{Cs}{} samples. Measured microwave cavity characteristics for the resonator loaded with the sample of Cs- or Rb-nabokoite smoothly change with temperature as shown in Figure~\ref{fig:quality} practically following the reference curve for the empty sample-holder.

However, accurate measurements performed with larger samples do reveal additional low-temperature features in dielectric response of nabokoites (see Figure~\ref{fig:lt}, note that microwave absorption drastically depends on whether the sample is placed to the beam of the high-frequency electric or to the beam of the high-frequency magnetic field). Because of  weakness of this absorption, we do not attempt to determine quantitatively the dielectric permittivity and simply plot the temperature dependences of microwave power transmitted through the cavity with the sample. For Cs-nabokoite we observe weak dielectric absorption at $T^*_\textrm{Cs}=(4.44\pm0.05)$~K, for Rb-nabokoite similar absorption is observed at $T^*_\textrm{Rb}=(4.65\pm0.05)$~K. For Na-nabokoite high-frequency dielectric absorption is observed  over the broader temperature range with maximum at $T^*_\textrm{Na}=(5.0\pm0.3)$~K and dielectric absorption is accompanied by high-frequency magnetic
absorption which tends to increase below the N\'{e}el point. As for the K-nabokoite, extended low-temperature wing of the main dielectric anomaly at 26~K prevents observation of such a weak features (however, weak high-frequency magnetic absorption was observed for K-nabokoite below the N\'{e}el point \cite{kolhoz} due to the opening of the spin waves gap). One can note (see inset to Figure~\ref{fig:lt}) that there is some change of slope on the experimental curve at $T^*_\textrm{K}=(6.5\pm1.0)$~K, but this feature can not be well resolved.

Amplitude of the low temperature dielectric losses can be estimated via nonlinear relation between the microwave power transmitted and the microwave power absorbed be the sample \cite{pool,bog}: $P_\textrm{transm}\propto 1/(1+W_\textrm{abs}/W_0)^2$ (here $W_\textrm{abs}$ is the power absorbed by the sample and $W_0$ is the power loss in the empty resonator due to Ohmic or other dissipation channels). This yields that the low-temperature dielectric absorption observed in Na-, Rb- and Cs-nabokoites corresponds to approximately 1/500 (per unit mass) of the absorption observed at the main dielectric anomaly in K-nabokoite at 26~K.

\section{Discussion}
\begin{figure*}[p]
 \centering
\includegraphics[width=0.6\textwidth]{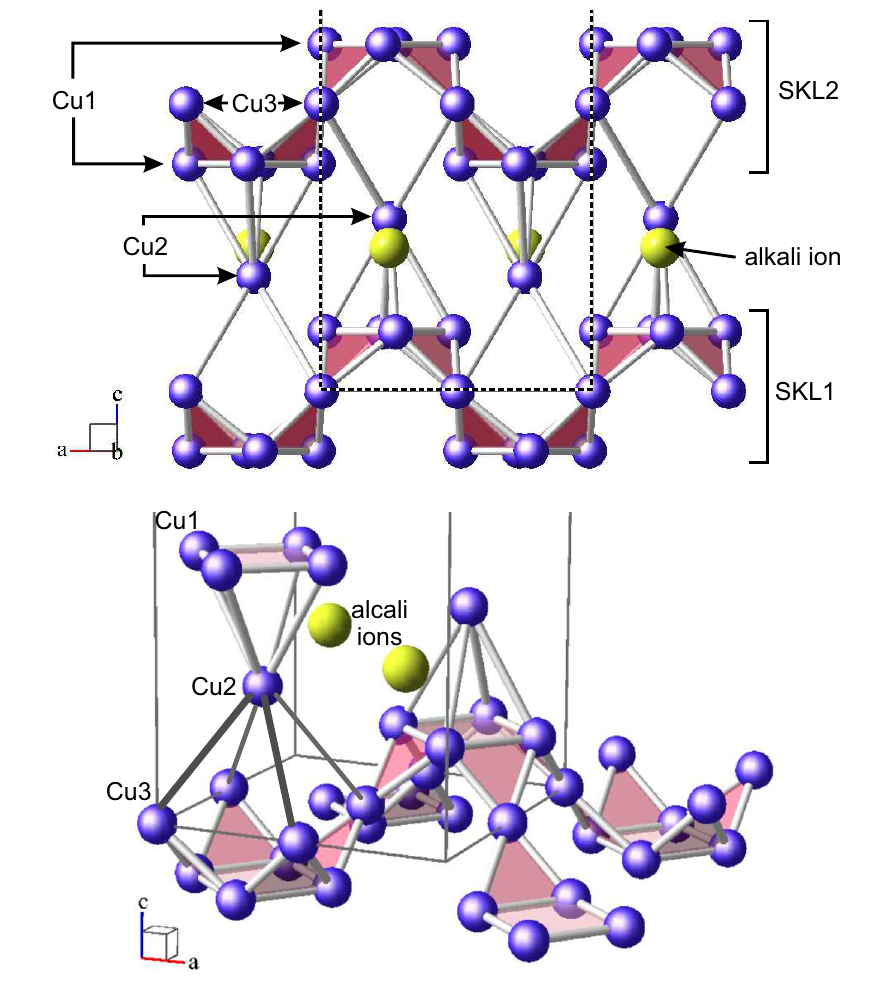}
 \caption{Fragments of nabokoite \nabok{A}{} crystallographic structure. Only positions of copper and alkali ions are shown. Atomic positions from \cite{Murtazoev2023}. Upper panel: View along the $b$-axis. 'SKL1' and 'SKL'2 are two SKL layers formed by Cu$^{2+}$ ions in 'Cu1' and 'Cu3' positions, Cu$^{2+}$ ions in 'Cu2' position decorates square \kagome{} layers and provides bridges for possible interlayer coupling. Dashed lines show unit cell boundaries. Lower panel: Isometric view of crystallographic structure fragment clarifying positions of decorating 'Cu2' ions and alkali ions with respect to 'Cu1' and 'Cu3' ions in neighboring square \kagome{} layers. Thin grey lines are unit cell boundaries.}
 \label{fig:struct}
\end{figure*}

\begin{figure*}[th]
 \centering
\includegraphics[width=\textwidth]{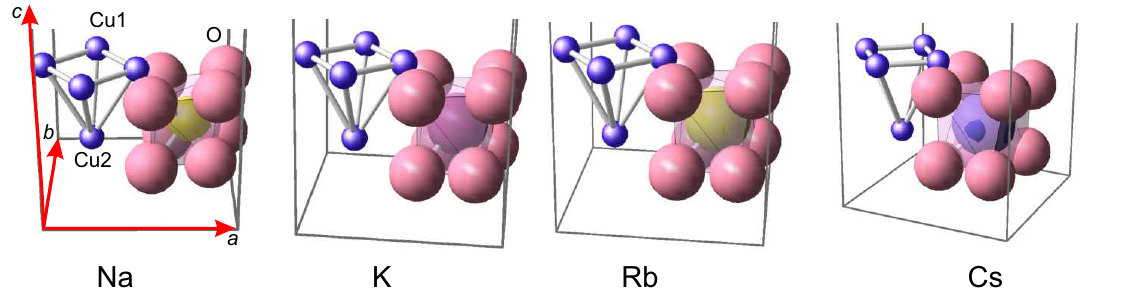}
 \caption{Structure fragments for \nabok{A}{} family. Positions of Cu1 ions (squares in SKL layer) and Cu2 ions (inter-layer copper ions), and positions of alkali ion with nearest 8 oxygens (O1 and O3) are shown. Atomic positions from \cite{Murtazoev2023}. Ionic radii \cite{Shannon1976} are shown up to the scale.}
 \label{fig:cages}
\end{figure*}

\begin{figure*}[th]
 \centering
\includegraphics[width=\textwidth]{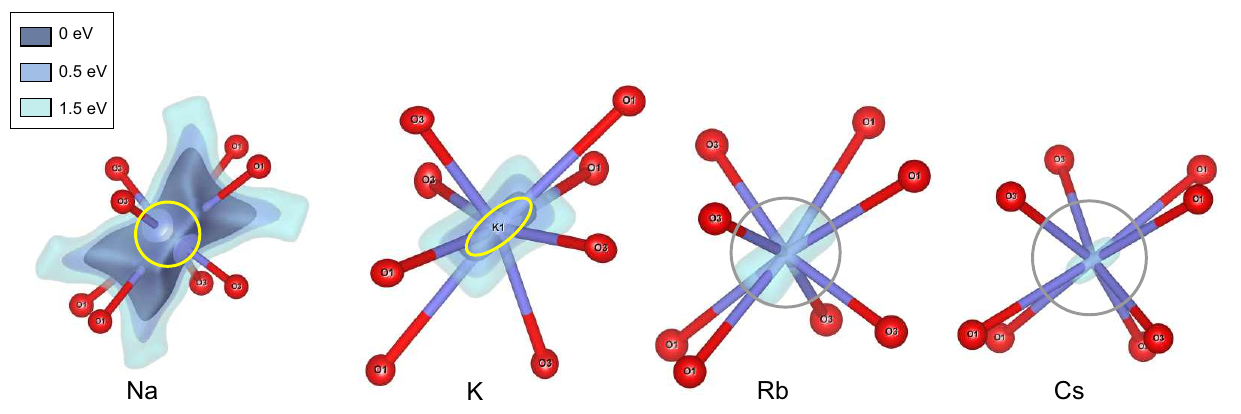}
\caption{Bond-valence site energy iso-surfaces for alkali ions in nabokoite family compounds \nabok{A}{} (A=Na, K, Rb, Cs). Energy iso-surfaces are calculated for $E=0, 0.5, 1.5$~eV (see legend). Circles on the panels drawn for Na, Rb and Cs compounds have a radius equal to ionic radius of alkali ion. Ellipse on the panel drawn for K compound shows thermal ellipsoid determined at 100~K \cite{Murtazoev2023}.}
 \label{fig:iso}
\end{figure*}

\begin{table*}[th]
    \caption{Summary of relevant information  for nabokoite family compounds \nabok{A}{}: characteristic temperatures(known antiferromagnetic Curie-Weiss temperature, N\'{e}el temperature or magnetic susceptibility $\chi(T)$ anomaly temperature, position of the low temperature specific heat $C(T)$ hump, position of static dielectric permittivity $\varepsilon(T)$ anomaly as well as temperatures of high-frequency dielectric anomalies in real and imaginary parts of the dielectric permittivity determined in the present work); parameters of alkali ion and its surroundings (alkali ions atomic masses $M_\textrm{alkali}$ and ionic radii $r_\textrm{alkali}$ (assuming 8-fold coordination \cite{Shannon1976}), volumes of nearest 8-vertices oxygen cage $V_8$ (calculated from structure data \cite{Murtazoev2023}), ratio of ion volume to the oxygen cage volume $V_\textrm{ion}/V_8$); lengths of the interlayer Cu-Cu bonds (calculated from the structural data of \cite{Murtazoev2023}). Question marks (?) indicate badly identified features (broad peaks and alike).}
    \centering
\begin{ruledtabular}    
\begin{tabular}{ ccccc }
    
     A=&Na & K & Rb & Cs\\
     &&&&\\
     \hline
     \multicolumn{5}{l}{Static data (from \cite{Murtazoev2023,kolhoz}):}\\
     $\Theta$, K &$>200$&$>200$&$>200$&80\\
     $\chi(T)$ features, K&3.8 ($T_\textrm{N}$)&3.2($T_\textrm{N}$)& 4.8 (?)& 4.3 (glass?)\\
     $C(T)$ hump, K&5.5&5.5&5.5&5.5 (broad)\\
     static $\varepsilon$ peak, K&--&25.4&--&--\\
     \hline
     &&&&\\
     \multicolumn{5}{l}{High-frequency dielectric anomalies (present work):}\\
     $T_{\varepsilon'}$, K & $100\pm5$ & $26.5\pm0.5$ & -- & --\\
     $T_{\varepsilon''}$, K& $90\pm10$ & $24\pm1$ & -- & --\\
     low-temperature $T^*$, K&$5.0\pm0.3$&$6.5\pm1$ (?)&$4.65\pm0.05$&$4.44\pm0.05$\\
     \hline
     &&&&\\
     \multicolumn{5}{l}{Structural information (calculated from data of \cite{Murtazoev2023}):}\\
     $M_\textrm{alkali} $,\text{ g/mole} & 23  & 39 & 85.5 & 133\\
     $r_\textrm{alkali} $, \r{A} \cite{Shannon1976} & 1.18 & 1.51 & 1.61 & 1.74\\
     $V_8$, \AA$^3$ &32.94&35.81&39.30&42.33\\
     $V_\textrm{ion}/V_8$&0.21&0.40&0.44&0.52\\
     Cu1-Cu2, \AA &4.533&4.629&4.700&4.843\\
     Cu2-Cu3, \AA &5.286&5.362&5.408&5.410\\
    \end{tabular}
\end{ruledtabular}
    \label{tab:disc}
\end{table*}

Our experiments demonstrate presence of the strong high-frequency dielectric anomalies in \nabok{Na}{} and \nabok{K}{}. Static dielectric anomaly was earlier reported for K-nabokoite at approximately the same temperature $T_\textrm{AFE}= 25.4$~K \cite{kolhoz}. Specific heat $C(T)$ measurements of Ref.\cite{kolhoz} does not demonstrate any sharp features at the temperature of dielectric anomaly. However, subtraction of high-temperature lattice contribution approximated by Debye-Einstein model reveals a broad peak in the $C(T)$ curve in the vicinity of the observed dielectric anomaly \cite{kolhoz}. No similar data for Na-nabokoite are available.

Quite small (two-fold) increase of the static dielectric permittivity in potassium nabokoite at this point \cite{kolhoz} suggests that this transition is the antiferroelectric one \cite{kittel-AFE}. Antiferroelectric materials are relatively rare, however they are known as useful medias for high-performance capacitors or microwave band-path filters \cite{AFE-review}. Effect of the applied electric field on dielectric properties of nabokoites is, by now, a separate challenge, hindered by the absence of single-crystalline samples. Observation of the strong temperature anomaly in dielectric losses ($\varepsilon''$) for potassium and sodium nabokoites (Figure~\ref{fig:eps'-eps''}) suggests possible application of these compounds as temperature-controlled broad-band microwave absorbers.

Shape of the observed anomalies in Na- and K-nabokoites is asymmetric, peaks of $\varepsilon'(T)$ and $\varepsilon''(T)$ are quite broad and slightly differ in position, the losses peak  being observed slightly below the peak temperature for the real part of dielectric permittivity. Broad and asymmetric peaks in dielectric constant are not unheard of for antiferroelectrics, see e.g. \cite{AFE-example}. We can not exclude that the peaks' shape is affected by polycrystalline form of the samples, as grain size is known to affect antiferroelectric transition \cite{AFE-review}.

We will focus our discussion here on the fundamental problem of magnetic ordering observed in Na- and K- compounds at low temperatures. Magnetic subsystem of nabokoites \nabok{A}{} is a unique decorated square \kagome{} lattice  (Figure~\ref{fig:struct}) \cite{Murtazoev2023,kolhoz}. Nabokoite formula unit includes seven magnetic Cu$^{2+}$ ions. Six copper ions form 2D SKL layers (four ions in 'Cu1' crystallographic position and two ions in 'Cu3' crystallographic position), while the seventh copper ion occupies inter-layer 'Cu2' positions. This yields two different frustration motives: (i) 2D SKL itself is a frustrated lattice, and (ii) decorating copper ion is located in the apex of square-based pyramid being equally coupled to four in-layer copper ions. The later case results in the effective decoupling of the apical ion from antiferromagnetically correlated basal plane ions within the classical mean field approximation. Large difference between the Curie-Weiss temperatures (80~K for Cs compound and above 200~K for other nabokoites \cite{Murtazoev2023,kolhoz}) and N\'{e}el temperatures (below 4~K, if any) indicates that frustration of exchange bonds does affect magnetic properties of nabokoites.

As one can see from Table~\ref{tab:disc}, presence of the main dielectric anomaly in Na- and K-nabokoites correlates with the low-temperature magnetic ordering. The microscopic origin of this anomaly remains unclear. However, we can draw some plausible hypotheses from the known structural data \cite{Murtazoev2023} (see also Supplementary Data for Ref.~\cite{Murtazoev2023}). Each alkali ion in \nabok{A} structure has 8 nearest oxygen ions, forming a polyhedra shown at Figure~\ref{fig:cages}. (Accurate description of the alkali ion coordination in nabokoites uses 12-vertices oxygen polyhedra \cite{Murtazoev2023}, but more simple 8-vertices oxygen cage is sufficient for our qualitative considerations.)  Volumes  of these 8-vertices oxygen cages ($V_8$) compared  with alkali ion ionic volume ($V_\textrm{ion}$) are given in Table~\ref{tab:disc}. One can see, that Na ion is quite unrestricted within the oxygen cage, while heavier ions are more and more constrained by oxygen surroundings (the largest Cs ion even ``pushes out'' through the cage faces, as seen at Figure~\ref{fig:cages}).

To make this discussion more quantitative we calculated bond-valence site energy iso-surfaces for the alkali ions in nabokoite family compounds using the SoftBV software package \cite{soft}. This software package uses Modern valence band theory by I.~D.~Brown (see, e.g., \cite{brown}). Calculated  iso-surfaces allow to visualize possible displacement of the alkali ions (see Figure~\ref{fig:iso}). One can see that for Na and K ions bond-valence site energy iso-surfaces has a pronounced butterfly-tie shape with characteristic length well exceeding ions' dimensions. The same modeling demonstrates that the heavy alkali ions (Rb and Cs) are tightly restricted at the nominal crystallographic position. This agrees with the structural analysis of \cite{Murtazoev2023} performed for K and Cs nabokoites (since only for these compounds it was possible to produce small single-crystalline pieces suitable for structure refinement). Anisotropic displacement parameters for K ions within the $(ab)$-plane exceed those for other ions and that for K-ions in the $c$-direction approximately five-fold and the resulting thermal ellipsoid is aligned along the butterfly-tie shaped bond-valence site energy iso-surface (Figure~\ref{fig:iso}). For the heaviest Cs ion in \nabok{Cs}{} no remarkable anisotropy of displacement parameters was observed neither in X-ray experiment of \cite{Murtazoev2023} nor in the present simulations.

These structural details allow to suggest that, probably, the lightest alkali ions (Na and K) can either move within the oxygen cage at high temperatures or to occupy random energy-equivalent positions, while at low temperatures alkali ions order, their displacement providing the antiferroelectric pattern of cell polarizations leading to observed dielectric anomalies.  Heavy alkali ions (Rb and Cs) are more tightly restricted by their surroundings which prevents ions displacement.

The minute displacement of alkali ion within the $(ab)$-plane can be the key to low-temperature magnetic ordering in Na- and K-nabokoites. Alkali ions positions in \nabok{A}{} structure are located between the SKL layers formed by copper ions occupying 'Cu1' and 'Cu3' positions, approximately at the same level as the inter-layer copper ions taking 'Cu2' crystallographic positions (see Figure~\ref{fig:struct}). As it is noted above, the equivalence of four Cu1-Cu2 bonds leads to effective decoupling of Cu2 ion from antiferromagnetically correlated Cu1 ions within the mean field approximation. This suppresses the inter-layer coupling and prevents formation of 3D magnetic ordering at low temperatures. Displacement of the alkali ion within the $(ab)$-plane could cause displacement of Cu2 ion (or displacement of Cu2 ion surrounding involved in the super-exchange path), thus violating the equivalence of Cu1-Cu2 exchange bonds, which could ultimately lead to the formation of antiferromagnetic ordering at low temperatures. One can also note that the increase of Cu1-Cu2 and Cu2-Cu3 bonds' length (Table~\ref{tab:disc}) with the size of alkali ion also suppresses the inter-layer exchange coupling and acts also in the favor of suppressing 3D long-range magnetic ordering in Rb- and Cs-nabokoites. Identification of the alkali ions displacements is a separate quest requiring accurate structural measurements at low temperatures or some sensitive indirect methods (e.g., $^{39}$K NMR). Electron spin resonance data of Ref.~\cite{kolhoz} suggest that lattice symmetry within the low-temperature antiferromagnetic phase in K-nabokoite remains tetragonal.

Finally, we will discuss the weak low-temperature dielectric absorption observed in Na-, Rb- and Cs-nabokoites at approximately 4.5~K (Figure~\ref{fig:lt}). There is no obvious fit between the observed dielectric absorption and the known low-temperature physical properties of nabokoites (see Table~\ref{tab:disc}). N\'{e}el transitions were reported  for potassium and sodium nabokoites at  3.2~K  and 3.8~K, correspondingly \cite{Murtazoev2023,kolhoz}. Magnetization measurements on Cs-nabokoite \cite{Murtazoev2023} suggest possible spin-glass-freezing-like behavior at 4.3~K, while magnetic susceptibility $\chi(T)$ curve for Rb compound revealed weak unidentified hump-like feature at 4.8~K. Low-temperature specific heat measurements \cite{kolhoz,Murtazoev2023} revealed hump-shaped anomalies for K, Rb and Na compounds around 5.5~K, which are not affected by applied magnetic field, much broader hump is observed for Cs compound around the same temperature. The low-temperature specific heat $C(T)$ hump for K-nabokoite was discussed as a possible evidence of the singlet states present within the triplet energy gap, as expected for ideal SKL magnet \cite{kolhoz}. Weak low-temperature dielectric anomalies are observed for Na-, Rb- and Cs-nabokoites (Figure~\ref{fig:lt}) close to these characteristic temperatures found in static measurements. This may indicate complex interplay of magnetic, dielectric and elastic properties of nabokoites at low temperatures, which is a subject of future studies by a more sensitive techniques. Concluding this part of discussion we note also, that the low temperature phases of K- and Na-nabokoites are,  supposedly, a rare example of multiferroic material combining \emph{antiferromagnetic} and \emph{antiferroelectric} orders.

\section{Conclusions}

We report results of the study of high-frequency ($\sim 10$~GHz) dielectric response of nabokoite-family compounds \nabok{A} (A=Na, K, Rb, Cs), which are the example of a quasy-2D square \kagome{} lattice (SKL) magnet with additional ``decorating'' inter-layer magnetic ions.

We have observed strong anomalies both in real and imaginary parts of the high-frequency dielectric permittivity for K- and Na-nabokoites at approx. 26 and 100~K, correspondingly. These anomalies are presumably related to antiferroelectric transition. Minute lattice deformations appearing at these transitions are probably the key ingredient to the formation of antiferromagnetic long-range order at much lower temperatures observed for these compounds.

Most of the family members (A=Na, Rb, Cs) demonstrate also much weaker dielectric absorption at low temperatures (4-6~K), which probably indicates complex interplay of magnetic and dielectric (lattice) degrees of freedom in nabokoites.

While the physics of the decorated square \kagome{} lattice magnets remains enigmatic and possible magnetically ordered low-temperature phases are not well understood, nabokoite family compounds provides interesting playground open both for theoretical and experimental efforts.

\section*{Acknowledgements}
The work was supported be Russian Science Foundation grants No. 22-12-00259 (high-frequency experiments) and No. 23-23-00205 (samples preparation and X-ray characterization). Some of the authors (Ya.V.R. and V.N.G.) acknowledge support  within the framework of the Academic Fund Program at HSE University via grant No. 24-00-011 (data processing).

\appendix

\section{Details of the high-frequency technique \label{sec:app}}
\begin{figure*}[th]
 \centering
\includegraphics[width=0.8\textwidth]{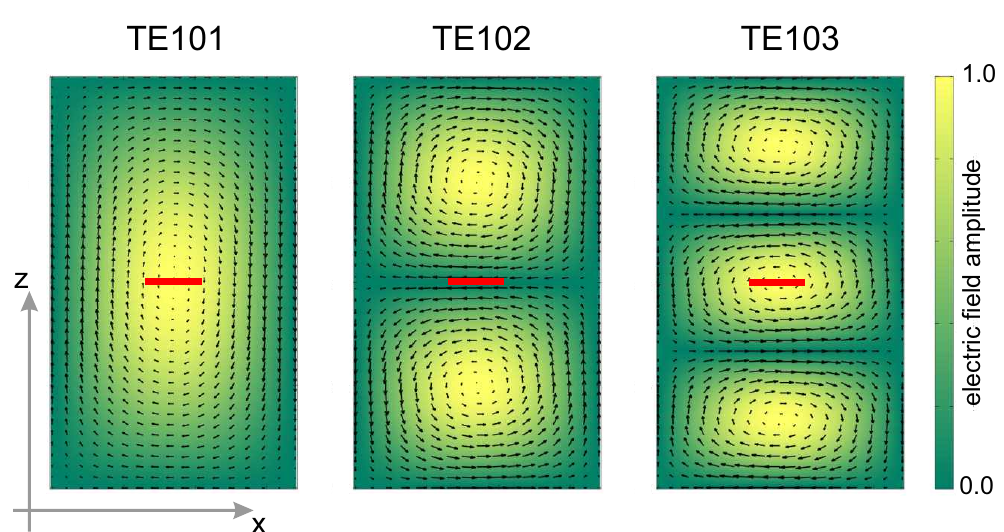}
\caption{High-frequency fields distribution for different TE$10k$ eigenmodes of the rectangular cavity used in the present study. Electric field is directed along $y$ axis (normal to the figure plane), its amplitude $\abs{E_y}$ is shown by the color map. Vectors represent directions and amplitudes of the magnetic field. All fields are uniform along $y$ axis. Color bar at the cavity center represents relative size of the sample.}
 \label{fig:fields}
\end{figure*}

Eigenmodes of the rectangular resonator \cite{jackson,pool,bog} are the standing waves of high-frequency electromagnetic field. They are conventionally denoted as TE$mnk$, here $m$, $n$ and $k$ are integer numbers of half-waves along $x$, $y$, $z$ directions (see Figure \ref{fig:setup} for the choice of the axes). Lowermost eigenfrequencies of our cavity corresponds to $n=0$ (i.e. uniform electromagnetic field along the shortest dimension of the cavity), and $m=1$ (one half-wave along the middle dimension of the cavity). Amplitudes of the high-frequency fields for TE$10k$ modes are described by equations:
\begin{eqnarray}
E_y&=&A \sin(\pi x/L_x) \sin(k\pi z/L_z)\\
H_x&=&\imath A\frac{k \pi/L_z}{K} \sin(\pi x/L_x)\cos(k \pi z/L_z)\\
H_z&=&-\imath A\frac{\pi/L_x}{K}\cos(\pi x/L_x) \sin(k\pi z/L_z)
\end{eqnarray}

\noindent here $K=\sqrt{\left(\frac{\pi}{L_x}\right)^2+\left(\frac{k\pi}{L_z}\right)^2}$, $L_{x,y,z}$ are the cavity dimensions, all other field components are zero, complex $\imath$ multiplier for magnetic field components is due to the phase shift between oscillations of electric and magnetic fields.

Distributions of the electromagnetic fields in the cavity modes used in our experiments are shown in Figure~\ref{fig:fields}. Beams of the electric field coincide with the nodes of magnetic field and vice versa. This allows to place small sample controllably either to the beam of the electric field (TE101 and TE102 modes for our setup) or to the beam of the magnetic field (TE102 mode) and to control polarization of the high-frequency fields at the sample location. Our sample dimensions are small compared to cavity sizes allowing to consider electromagnetic fields at sample location as uniform.

Problem of effect of dielectric ellipsoid on microwave cavity properties is solved in some textbooks (e.g., \cite{bog,LL8}), it is also closely related to the known problem of calculation of magnetic resonance absorption effect on a microwave cavity properties \cite{pool}.

Shift of the cavity eigenfrequency under introduction of dielectric ellipsoid is \cite{bog,LL8}

\begin{equation}\label{eqn:app-domega}
\frac{\delta\omega}{\omega}=-2\pi \frac{\beta \left(\mathbf{E}\left(\mathbf{r}^{(\textrm{s})}\right)\right)^2}{\iiint_{V^{(\textrm{res})}}\left(\mathbf{E}(\mathbf{r})\right)^2 dV}
\end{equation}

\noindent here $\mathbf{E}\left(\mathbf{r}^{(\textrm{s})}\right)$ is the electric field at the sample location, $V^{(\textrm{res})}$ is the cavity volume, $\beta$ is the polarizability coefficient defining {\em total} electric polarization of the small sample ${\cal P}=\beta \mathbf{E}\left(\mathbf{r}^{(\textrm{s})}\right)$. For the thin disk with the field applied parallel to the plane (analogously to the  demagnetization factor calculations \cite{jackson,LL8}) polarizability is equal to $$\beta=\frac{\varepsilon'-1}{4\pi} V^{(\textrm{s})}$$
 here $V^{(\textrm{s})}$ is the sample volume.

Assuming that the sample is located exactly at the beam of the high-frequency electric field of TE$10k$ mode,  integral in the denominator of (\ref{eqn:app-domega}) yields
$$\iiint_{V^{(\textrm{res})}}\left(\mathbf{E}(\mathbf{r})\right)^2 dV=\frac{1}{4} \left(\mathbf{E}\left(\mathbf{r}^{(\textrm{s})}\right)\right)^2 V^{(\textrm{res})}$$
factor 1/4 here is due to the averaging of $\mathbf{E}^2$ over half-periods of spacial oscillations along $x$ and $z$ directions.

This ends up with the final equation on the real part of permittivity

\begin{equation}
\varepsilon'=1 - \frac{V^{(\textrm{res})}}{2V^{(\textrm{s})}}\cdot \frac{\delta \omega}{\omega_0}
\end{equation}

Imaginary part of the dielectric permittivity causes diamagnetic losses with time-averaged dissipated power $$W^{(\textrm{diel})}=\frac{\omega\varepsilon''}{8\pi}\left(\mathbf{E}\left(\mathbf{r}^{(\textrm{s})}\right)\right)^2 V^{(\textrm{s})}$$
Cavity Q-factor is defined as
$$Q=\frac{\omega E^{(\textrm{stored})}}{\sum W_i}$$
here $E^{(\textrm{stored})}$ is the energy stored in the cavity and the total losses power $\sum W_i$ is the sum over all dissipation channels. The difference of the inverse Q-factors directly yields contribution of dielectric losses:
\begin{equation}
\frac{1}{Q^{(\textrm{s})}}-\frac{1}{Q^{(\textrm{e})}}=\frac{W^{(\textrm{diel})}}{\omega E^{(\textrm{stored})}}
\end{equation}

\noindent here $Q^{(\textrm{s})}$ and $Q^{(\textrm{e})}$ are Q-factors for the cavity with the sample and for the cavity with empty sample-holder.

Assuming that the sample is positioned at the beam of the electric field one can calculate energy stored in the cavity for TE$10k$ modes
\begin{eqnarray}
E^{(\textrm{stored})}&=&\frac{1}{2}\iiint_{V^{(\textrm{res})}}\left(\frac{\mathbf{H}^2}{8\pi}+\frac{\mathbf{E}^2}{8\pi}\right) dV=\nonumber\\
&=&\frac{1}{8\pi}\times \frac{1}{4}\times \left(\mathbf{E}\left(\mathbf{r}^{(\textrm{s})}\right)\right)^2 V^{(\textrm{res})}\nonumber
\end{eqnarray}
here factor 1/2 at the first integral is the result of the time-averaging of the oscillating fields and factor 1/4 at the last expression is the result of $\mathbf{E}^2$ averaging over half-periods of spacial oscillations along $x$ and $z$ directions.

This yields expression for the imaginary part of dielectric permittivity

\begin{equation}
\varepsilon''=\frac{1}{4} \frac{V^{(\textrm{res})} }{V^{(\textrm{s})}} \left(\frac{1}{Q^{(\textrm{s})}}-\frac{1}{Q^{(\textrm{e})}}\right)
\end{equation}


\end{document}